\documentclass[conference]{IEEEtran}
\usepackage{graphicx}
\usepackage{caption}
\usepackage{subcaption}

\usepackage{longtable}
\usepackage{stfloats}
\usepackage{graphicx}

\usepackage{booktabs}
\usepackage{multirow}

\usepackage{ifpdf}
\usepackage[T1]{fontenc}
\usepackage{amssymb}
\usepackage{cite}

\ifCLASSINFOpdf
   \usepackage{graphicx}
	\usepackage{epstopdf}
   \graphicspath{{images/}}   
   \graphicspath{{../pdf/}{../jpeg/}{../eps/}}
   \DeclareGraphicsExtensions{.pdf,.jpeg,.png,.eps}
\else
   \usepackage[dvipdfm]{graphicx}
   \usepackage{bmpsize}
   \graphicspath{{../eps/}}
   \DeclareGraphicsExtensions{.eps}
\fi
\graphicspath{{images/}}

\usepackage[cmex10]{amsmath}
\usepackage{multirow}

\usepackage{multirow}

\usepackage{booktabs}
\usepackage{multicol}

\usepackage[shortcuts,acronym, nomain]{glossaries}

\makeglossaries 

\newacronym{rnn}{RNN}{Recurrent Neural Network}
\newacronym{awgn}{AWGN}{Additive White Gaussian Noise}
\newacronym{gru}{GRU}{Gated Recurrent Unit}
\newacronym{lstm}{LSTM}{Long-Short Term Memory}
\newacronym{snr}{SNR}{Signal to Noise Ratio}
\newacronym{urllc}{URLLC}{Ultra Reliable Low Latency Communication}
\newacronym{dl}{DL}{Deep Learning}
\newacronym{ml}{ML}{Machine Learning}
\newacronym{qos}{QoS}{Quality of Service}
\newacronym{mmtc}{MMTC}{Massive Machine Type Communications}
\newacronym{iot}{IoT}{Internet of Things}
\newacronym{cnn}{CNN}{Convolutional Neural Network}
\newacronym{gan}{GAN}{Generative Adversarial Network}
\newacronym{dbn}{DBN}{Deep Belief Network}
\newacronym{phy}{PHY}{Physical Layer}
\newacronym{ann}{ANN}{Artificial Neural Network}
\newacronym{mlp}{MLP}{Multi Layer Perceptron}
\newacronym{ldpc}{LDPC}{Low-Density Parity-Check}
\newacronym{ber}{BER}{Bit Error Rate}
\newacronym{gpu}{GPU}{Graphical Processing Unit}
\newacronym{rsc}{RSC}{Recursive Systematic Convolutional}
\newacronym{map}{MAP}{Maximum A Posteriori}
\newacronym{adam}{ADAM}{Adaptive Moment Estimation}
\newacronym{qpsk}{QPSK}{Quadrature Phase Shift Keying}
\newacronym{qam}{QAM}{Quadrature Amplitude Modulation}

\newacronym{ofdm}{OFDM}{Orthogonal Frequency Division Multiplexing}
\newacronym{csi}{CSI}{Channel State Information}
\newacronym{ls}{LS}{Least-Squares}
\newacronym{mmse}{MMSE}{Minimum Mean Squared Error}
\newacronym{dmrs}{DMRS}{Demodulation Reference Symbols}
\newacronym{3gpp}{3GPP}{3rd Generation Partnership Project}
\newacronym{enb}{eNB}{eNodeB}
\newacronym{lmmse}{LMMSE}{Linear MMSE}
\newacronym{immse}{IMMSE}{Inverse MMSE}
\newacronym{nn}{NN}{Neural Network}
\newacronym{cv2x}{C-V2X}{Cellular-Vehicle-to-Everything}
\newacronym{prb}{PRB}{Physical Resource Block}
\newacronym{bler}{BLER}{Block Error Rate}
\newacronym{evm}{EVM}{Error Vector Magnitude}
\newacronym{mse}{MSE}{Mean Squared Error}
\newacronym{sc-fdma}{SC-FDMA}{Single Carrier - Frequency Division Multiple Access}

\begin{document}

\title{Channel Estimation in C-V2X using Deep Learning}
%
\author{\IEEEauthorblockN{Raja Sattiraju, Andreas Weinand and Hans D. Schotten}
\IEEEauthorblockA{Chair for Wireless Communication \& Navigation \\
University of Kaiserslautern\\
\{sattiraju, weinand, schotten\}@eit.uni-kl.de}}
\maketitle

\begin{abstract}
	Channel estimation forms one of the central component in current \ac{ofdm} systems that aims to eliminate the inter-symbol interference by calculating the \ac{csi} using the pilot symbols and interpolating them across the entire time-frequency grid. It is also one of the most researched field in the \ac{phy} with \ac{ls} and \ac{mmse} being the two most used methods. In this work, we investigate the performance of deep neural network architecture based on \acp{cnn} for channel estimation in vehicular environments used in 3GPP Rel.14 \ac{cv2x} technology. To this end, we compare the performance of the proposed \ac{dl} architectures to the legacy \ac{ls} channel estimation currently employed in C-V2X. Initial investigations prove that the proposed \ac{dl} architecture outperform the legacy C-V2X channel estimation methods especially at high mobile speeds.

\end{abstract}

\section{Introduction}

Starting with 3G, \ac{ofdm} has been the choice of \ac{phy} technology due to its resilience to inter-symbol interference and multipath fading, higher data rates and better spectrum efficiency. At the heart of any \ac{ofdm} receiver is the channel estimation function which estimates the \ac{csi} and uses this information to equalize the received waveform in order to mitigate the channel effects. Channel estimation can be performed in either time domain or frequency domain with the latter being used extensively in current \ac{ofdm} systems due to its simplicity and ease of implementation.\

In frequency domain channel estimation, known symbols called pilots are transmitted at known positions in the \ac{ofdm} resource grid. These pilots can be arranged in a few frequency sub-carriers in all \ac{ofdm} symbols (comb configuration), or across all subcarriers in few \ac{ofdm} symbols (block configuration) or across few subcarriers on few \ac{ofdm} symbols (2D grid configuration). At the receiver side, the channel is estimated by means of comparing the received pilot symbols with the transmitted pilot symbols thereby yielding the impulse response of the channel at these locations. By means of averaging and interpolation, these impulse responses are fine tuned to get the channel impulse response matrix $H$ of the whole transmitted \ac{ofdm} resource grid.\

There are primarily two methods for channel estimation \cite{Hermansson2011,  Beek1995}. The first one is based on the computationally simple \ac{ls} algorithm which directly divides the received pilot symbols with the transmitted pilots symbols in the frequency domain. However, this method ignores the effect of \ac{awgn} to which the \ac{ofdm} systems are very sensitive to. In order to reduce the effect of noise, the channel impulse responses after \ac{ls} estimation are averaged across time and frequency with a given 2D window size. Hence the noise is also averaged reducing its overall effect. The second method is the \ac{mmse} method that uses the statistical characteristics of the noise and the channel matrix; therefore, its computational complexity is high. Other methods such as \ac{lmmse}, \ac{immse}, Chi-square distribution-based method, Haar Wavelet based method were also proposed and basically depend on \ac{ls} and \ac{mmse} based methods albeit with higher complexity.\ 


 
Recently, \ac{ml}, which has shown substantial promises during recent years\cite{Goodfellow} is extensively studied by researchers in order to assess its applicability to wireless technologies.  \ac{dl} architectures such as \acp{cnn}, \acp{rnn}, \acp{gan}, \acp{dbn} have been applied to various domains such as computer vision, natural language processing, social network filtering, drug design etc. where they have produced results comparable to and in some cases superior to human experts.

In this work, we investigate \ac{dl} architectures to design and analyse an \ac{ann} for channel estimation in high mobility vehicular scenarios. Specifically, we use \ac{cv2x} as underlying technology that specifies the use of \ac{ls} based channel estimation. To this end, we frame the channel estimation as a supervised learning problem and use \ac{dl} architectures based on \ac{cnn} to output the channel estimation matrix $H$ which in turn is used for equalization.

\subsection{Related State of the Art}
Initially applied to upper layers \cite{Jiang2017}, \ac{ml} has recently found applications also at the \ac{phy} Layer \cite{Wang2017, Kim2018, OShea2017} such as channel coding\cite{Ortuno1992,Bruck1989, Sattiraju2018}, modulation recognition\cite{OShea2017}, obstacle detection \cite{Sattiraju2017a, Sattiraju2019} and physical layer security \cite{Weinand2017a} etc. The use of \ac{ml} for channel estimation was initially investigated in works such as \cite{Chen1990}. Subsequently, with the advent of programmable \acp{gpu} and availability of open source \ac{dl} libraries such as Tensorflow and caffe, many other works started applying various \ac{dl} architectures for channel estimation and equalization. In \cite{Sun2006}, the authors proposed an architecture by stacking two independent \acp{ann} on top of each other for estimating the amplitude and phase values directly. The use of \acp{cnn} for channel estimation was investigated in \cite{Balevi2019, Ye2018}. In \cite{Cheng2016}, the authors combined a \ac{ann} architecture with a genetic algorithm for channel estimation. In \cite{Soltani}, the authors used a combination of Super-resolution \ac{cnn} and a feed-forward de-noising \ac{cnn} to achieve performances close to \ac{mmse} methods. The use of multi-layer perceptron for channel equalization is investigated in \cite{Chen1990}.\

Our work differs from the previous work with respect to two major points. We use \ac{dl} architecture based on \ac{cnn} to directly output the channel matrix $H$ over different \ac{snr} points (as opposed to single \ac{snr} value used in other works) that can be readily used for equalization. Furthermore, we use the C-V2X sidelink as underlying technology in contrast to Uu based LTE where the pilot arrangements differ and the use of \ac{mmse} is not encouraged due to high mobility scenarios. 

The rest of the paper is organized as follows. Section II gives a broad overview about the \ac{cv2x} technology and the channel estimation method used. Section III outlines the simulation method using the proposed \ac{dl} architecture. It also outlines the training operation and the results. Section IV concludes the paper with some future directions

\section{Channel Estimation in \ac{cv2x}}
\ac{cv2x} has been proposed in order to enable direct communications in 3GPP Rel.14. It introduces a new interface called PC5 in addition to the legacy Uu interface in LTE to support direct communication between devices with or without the presence of an
\ac{enb}. In Rel.15, it was further enhanced to support V2X applications by increasing the  \ac{dmrs} symbols fro 3 to 4 to better tackle the fast channel variations in vehicular scenarios. V2X communications happen in periodic intervals (called the sidelink period) that ranges between 40 ms to 320 ms corresponding to 40-320 subframes. Any vehicle can transmit twice (with 1 blind retransmission) on any two selected subframes in time domain within this period (sidelink subframes). Within every sidelink subframe, there are resource pools allocated for sidelink transmission and the vehicle dynamically selects a subset of these resource pools (\acp{prb}) for transmission. In mode 3, the \ac{enb} controls the selection of sidelink subframes and resource pools whereas in mode 4, the vehicle autonomously select from a set of pre-configured resource pools.\

Each sidelink subframe (1 ms) contains 14 \ac{ofdm} symbols out of which 10 are used for carrying user data and the remaining 4 (at positions [2,5,8,11] with 0-based indexing) are used for carrying \ac{dmrs} symbols. The \ac{dmrs} symbols are sequences $r_{u,v}^{(\alpha)}$ that are obtained by a cyclic shift of a base sequence ${r}_{u,v}(n)$ according to 

\[ r_{u,v}^{(\alpha)} = e^{j\alpha n} \cdot \bar{r}_{u,v}(n),0\leq n \leq  M_{sc}^{RS} \]

where $M_{sc}^{RS} = mN_{sc}^{RB}$ is the length of the \ac{dmrs} sequence, $m$ is the number of \acp{prb} and $N_{sc}^{RB}$ is the number of subcarriers within one \ac{prb} (12 in case of LTE). The base sequence itself is defined as the cyclic extension of the Zadoff-Chu Sequence and is given as
\[ \bar{r}_{u,v}(n) = x_q\left ( n \mod N_{ZC}^{RS} \right ),0\leq n \leq M_{sc}^{RS} \]

\[ x_q(m) = e^{-j\frac{\pi q m (m+1)}{N_{ZC}^{RS}}},0 \leq m \leq N_{ZC}^{RS} - 1 \]

where $x_q(m)$ is the $q_{th}$ root of Zadoff-Chu sequence and $N_{ZC}^{RS}$ is the length of Zadoff-Chu sequence that is given by the largest prime number such that $N_{ZC}^{RS} < M_{sc}^{RS} < 3N_{sc}^{RB}$, the base sequence is defined as the computer generated constant amplitude zero autocorrelation (CG-CAZAC) sequence.

\[ \bar{r}_{u,v}(n) = e^{j\varphi (n)\pi /4},0 \leq n \leq M_{sc}^{RS} \]

The transmitting node can select a base sequence from a set of groups each differentiated with a hopping sequence that depends on the current subframe number and the V2X scrambling identity. In this way, the \ac{dmrs} sequences are randomized for different vehicles thereby reducing inter-cell interference.

The \ac{dmrs} symbols along with the data symbols are multiplexed together, modulated by \gls{sc-fdma} and then passed through a channel. At the receiver, the received \ac{ofdm} grid denoted as $Y_{t}$ is given as

\[ Y_{t} = H_{t}X_{t} + N_{t}, \]

where $H_{t}$ is the channel frequency response and $N_{t}$ is the \ac{awgn} for symbol $t$.

As a first step in \ac{ls} channel estimation, the receiver extracts the pilot symbols from their known location in the time-frequency grid and divides them with their expected value

\[ \tilde{H}_{(i,k)} = \frac{Y_{(i,k)}}{X_{(i,k)}} = H_{(i,k)} + N_{(i,k)} \]

where $\tilde{H}_{(i,k)}$ is the \ac{ls} channel estimate at pilot location $(i,k)$, $Y_{(i,k)}$ and $X_{(i,k)}$ are the received and sent pilot symbols at $(i,k)$ and $N_{(i,k)}$ is the noise at $(i,k)$. It can be seen that the calculated \ac{ls} estimate is noisy and hence in order to minimize the effect of noise, a 2D averaging is performed with a chosen window size. Hence, averaging the instantaneous channel estimates over the window, we have
		
\[ \tilde{H}^{AVG}(i,k) = \frac{1}{|S|}\sum_{m\in S}\tilde{H}_{(i,k)}(m) \approx H_{(i,k)}  \]

where $S$ is the set of pilots in the 2D window and $|S|$ is the number of pilots in $S$. The \ac{ls} estimates and the averaged estimates contain the same data, apart from additive noise. Simply taking the difference between the two estimates results in a noise level value for the \ac{ls} channel estimates at pilot symbol locations. This knowledge of noise can be useful to increase the performance of some receivers especially using soft demodulation techniques.

Finally, the averaged \ac{ls} estimates are interpolated across the whole time-frequency grid to get the complete channel matrix $H(t)$ for the received subframe. Equalization is performed by multiplying the received grid $Y(t)$ with the complex conjugate of $H(t)$

\[ Y^{eq}(t) = Y(t) * H(t)^* \]

\section{Simulation Methodology}

\begin{figure}[h!]
	\centering 
	\includegraphics[width=0.48\textwidth]{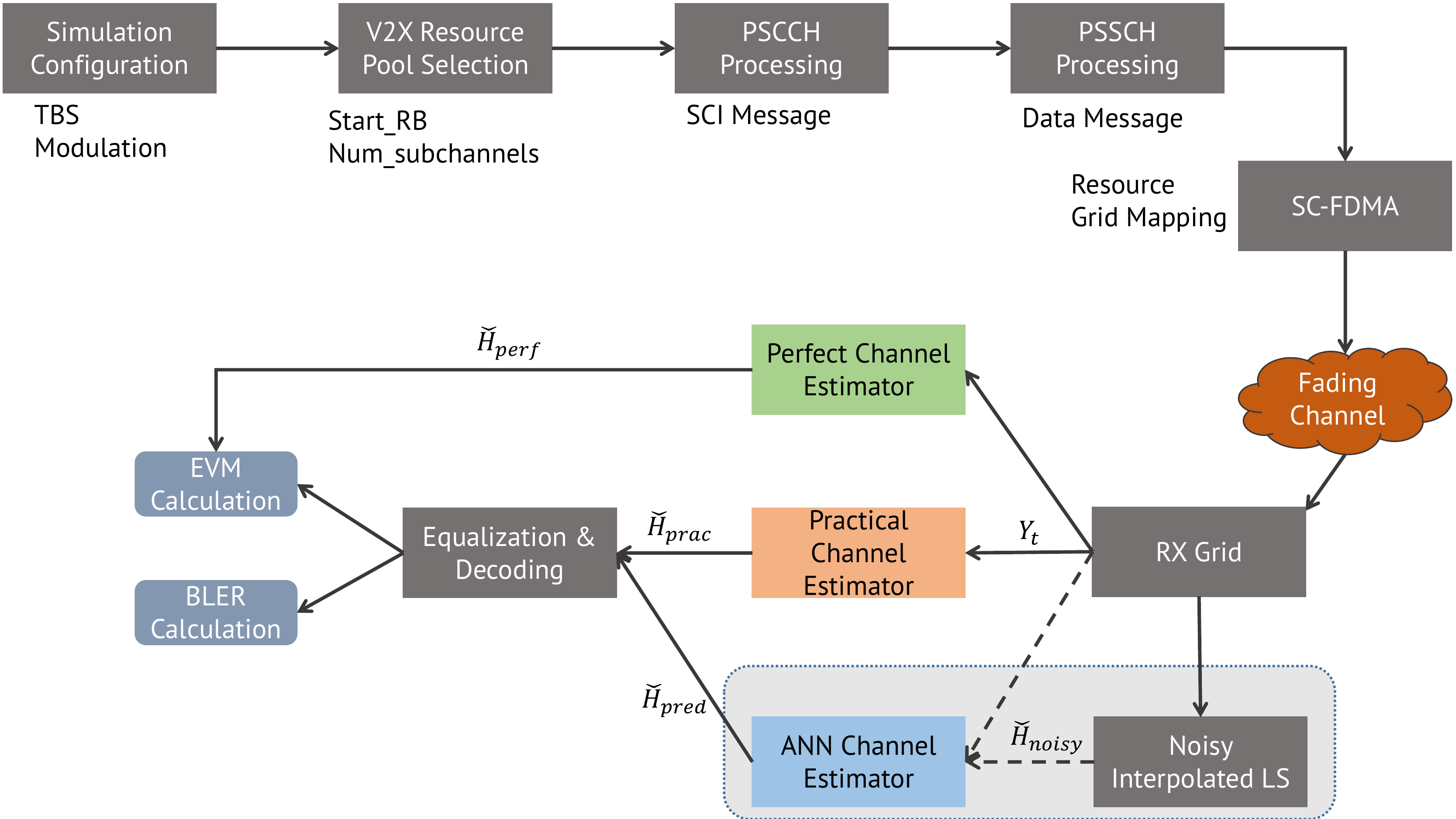}
	\caption{Simulation Method}
	\label{fig:simmethod}
\end{figure} 

As a proof-of-concept, we applied the \ac{ann} based channel estimation on simulated data. The simulation method is outlined in Figure.\ref{fig:simmethod}. For the given set of parameters as outlined in Table \ref{tab:sim_par}, we generated a set of sidelink subframes. These subframes were then converted to a time domain waveform by employing \gls{sc-fdma} and the waveform was passed through a multi-path fading channel (with \textit{EVA} delay profile) to get the received grid $Y_t$. The receiver operations consist of subframe synchronization followed by perfect and practical channel estimation that produced channel matrices ($\hat{H}_{perf}$) and $\hat{H}_{prac}$ respectively. The noisy \ac{ls} estimates were obtained by dividing the received \ac{dmrs} with the transmitted \ac{dmrs} symbols and this is linearly interpolated over each subframe to get $\hat{H}_{noisy}$.\

\begin{table}[]
	\centering
	\caption{Simulation Parameters}
	\label{tab:sim_par}
	\resizebox{0.48\textwidth}{!}{%
		\begin{tabular}{@{}lll@{}}
			\toprule
			Parameter Group & Name & Value \\ \midrule
			\multirow{5}{*}{High-level Parameters} & Bandwidth & 10 MHz \\
			& NSLRB & 48 \\
			& TBS & 3496 \\
			& N\_Subframes & 500 \\
			& SNR Range & {[}-2, 5{]} dB \\
			\multirow{2}{*}{SCI Message} & Modulation & QPSK \\
			& Time Gap & 1 subframe \\
			Data Message & Modulation & QPSK \\
			\multirow{3}{*}{Channel} & Delay Profile & EVA \\
			& MIMO & 1X2 \\
			& Speeds & {[}100,200,300,400{]} kmph \\ \cmidrule(l){2-3} 
		\end{tabular}%
	}
\end{table}

\subsection{Data Generation \& Preparation}
For the training data, we generated a total of 500 subframe samples for SNR values ranging between $[-2, 5]$ dB hence totaling 5000 samples. This process was repeated for 4 different speeds (Table.\ref{tab:sim_par}) bringing the total number of samples to 20000. Each sample is has a shape of $(576, 14)$ corresponding to 48 \acp{prb} in frequency domain and 14 symbols in time domain. $y$ also has a similar shape as $X$. 

\subsection{\ac{dl} architectures}
\acf{dl} belongs to a class of \ac{ml} algorithms that uses multiple layers of non linear processing units stacked on top of each other. Each successive layer uses the output of the previous layer as input. Such \ac{dl} architectures are especially suitable for designing auto encoders that aim to find a low-dimensional representation of its input at some intermediate layer that allows reconstruction at the output with minimal error. \ac{dl} architectures can be broadly classified into two categories - \acp{cnn} which are good at finding spatial patterns in the data and \acp{rnn} which are good at finding temporal correlations. \

\begin{figure}[h!]
	\centering 
	\includegraphics[width=0.48\textwidth]{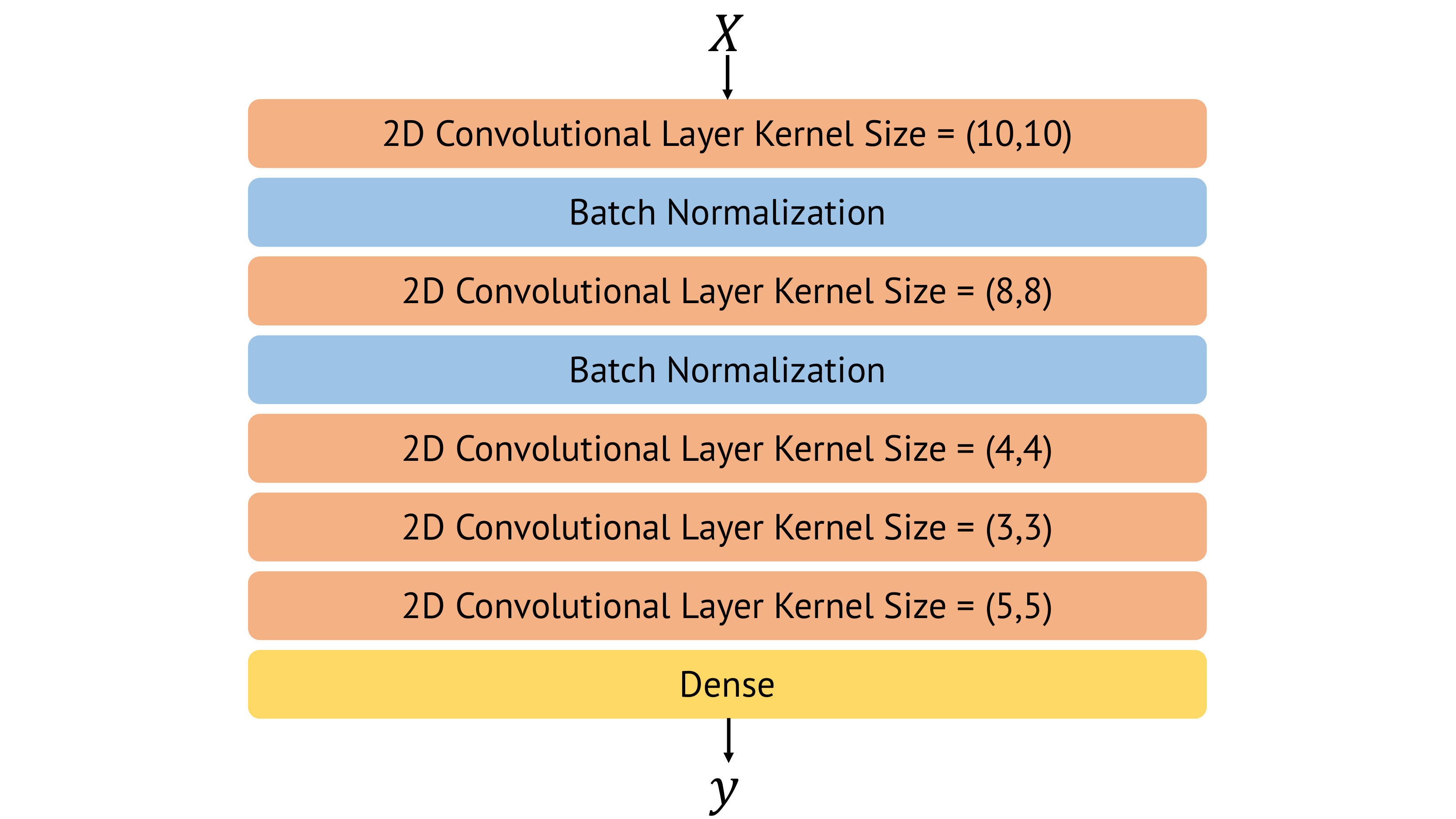}
	\caption{\ac{ann} Architecture}
	\label{fig:nn_arch}
\end{figure}

In this work, we adopt \ac{dl} architecture based on \ac{cnn} and compare their performance with respect to \ac{bler} and \ac{evm} to that of the perfect and practical channel estimators. As shown in Figure. \ref{fig:nn_arch}, the proposed model consists of 4 convolutional layers with different kernel sizes each of them followed by a batch normalization to minimize vanishing or exploding gradients. The final layer is a Dense layer followed by a reshape layer to reshape the data to have the same dimensions as the input data.\

\subsection{Training}
The input to the \ac{ann} is the noisy interpolated \ac{ls} channel matrix $\hat{H}_{noisy}$ and the output is the estimated channel matrix $\hat{H}_{pred}$

\[ \hat{H}_{pred} = f(\Phi;\hat{H}_{noisy}) \]
where $\Phi$ is the set of parameters of the \ac{ann}

For training, we used 30 \% of the samples. Finally, the trained model is used to output the $\hat{H}_{pred}$ for the whole sample set.

The loss function is the \ac{mse} between the estimated $\hat{H}_{est}$ and perfect channel matrix $\hat{H}_{perf}$ and is calculated as follows
\[ MSE = \frac{1}{\left \|\tau  \right \|}\sum_{h \in \tau}\left \| f(\Phi;\hat{H}_{pred})) - \hat{H}_{perf} \right \|^2 \]

For optimizing the loss, \ac{adam} optimizer was used. It computes the learning rates by calculating an exponentially decaying average of past gradients $m_t$ in addition to past squared gradients $v_t$ as follows\cite{Kingma2014}

\begin{equation*} 
\begin{split}
m_t & = \beta_1m_{t-1}+(1-\beta_1)g_t \\
v_t & = \beta_2v_{t-1}+(1-\beta_2)g^2_t
\end{split}
\end{equation*}

These values are then used to update the weights according to following rule
\[ \theta_{t+1} = \theta_t - \frac{\eta}{\sqrt{v_t} + \epsilon}m_t \]

We trained the network for 20 epochs and used this to predict $\hat{H}_{pred}$

\subsection{Evaluation}
The predicted channel $\hat{H}_{pred}$ is then used for equalizing the received grid. The equalized grid is then subsequently decoded and compared to the input data bits to obtain the \ac{bler}. In order to quantify the performance of the practical and \ac{ann} based channel estimator, the \ac{evm} was chosen as the metric that is calculated as follows

\[ EVM = \sqrt{\frac{\frac{1}{N}\sum_{k=1}^{N}(e_k)}{\frac{1}{N}\sum_{k=1}^{N}(I_k^2 + Q_k^2)}}*100 \]

where $e_k=(I_k-\hat{I}_k)^2 + (Q_k - \hat{Q}_k)^2$, $(I_k,Q_k) \& (\hat{I}_k, \hat{Q}_k)$ represent the In-phase component and the Quadrature phase component of the ideal and measured symbols respectively.

\subsection{Results}
\begin{figure*}[t!]
	\centering 
	\includegraphics[width=\textwidth]{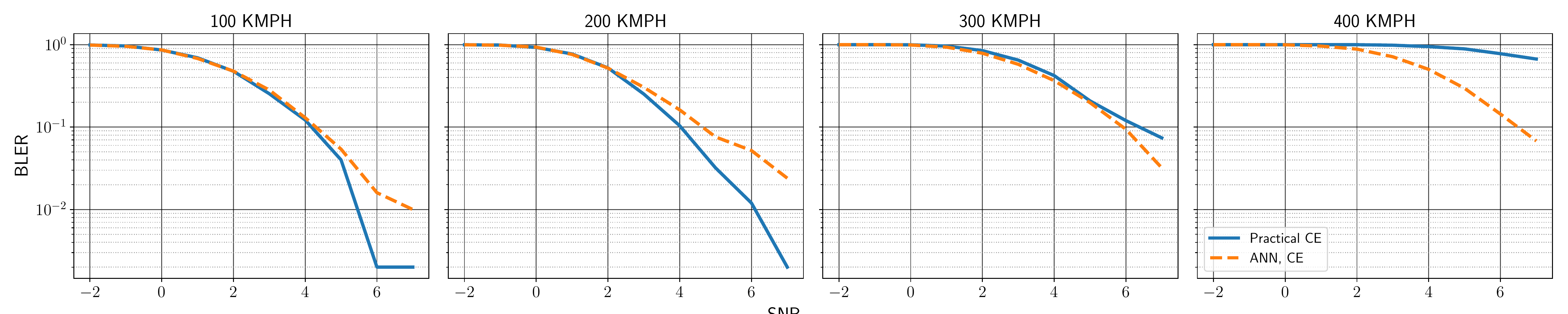}
	\caption{BLER Performance}
	\label{fig:results}
\end{figure*} 

\begin{figure*}[t!]
	\centering 
	\includegraphics[width=\textwidth]{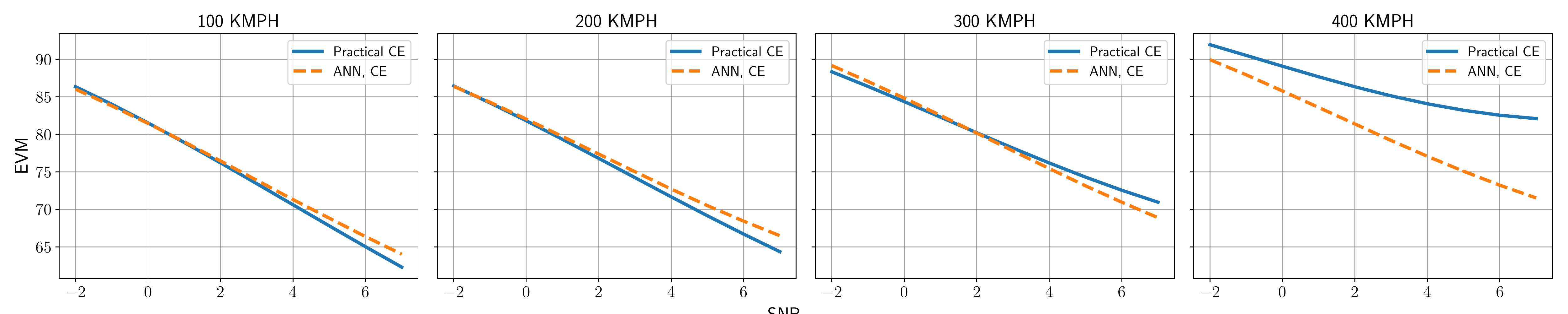}
	\caption{EVM Performance}
	\label{fig:results_evm}
\end{figure*} 

Figure \ref{fig:results} shows the \ac{bler} performance comparison between the practical channel estimator and the \ac{ann} based channel estimator. It can be clearly seen that the \ac{ann} based channel estimation scheme performs on par with the \ac{ls} scheme at low speeds and low SNRs. The real benefits of using \ac{ann} scheme become apparent at higher speeds and higher SNRs as the proposed scheme outperforms \ac{ls} scheme by almost an order of magnitude. This is because, at higher speeds, the averaging and interpolation used in \ac{ls} causes excessive information loss thereby resulting in pure noise. In contrast, \ac{ann} was better able to learn the quick channel variations in high speed scenarios. Figure \ref{fig:results_evm} also shows the \ac{evm} performance between the \ac{ann} and the \ac{ls} schemes. It can be seen that the \ac{evm} is almost identical for both the schemes at low speeds. At higher speeds and higher SNRs, \ac{ann} scheme shows lower \ac{evm} than the \ac{ls} scheme due to the effectiveness of the channel estimation.\

\section{Conclusions \& Future Work}
In this work, the use of \ac{dl} based on \ac{cnn} was investigated for the purpose of channel estimation in high mobility C-V2X scenarios. The proposed models were trained on data generated by means of simulations using the vehicular channel model with EVA delay profile. The trained models were used to output the predicted channel matrix. The BLER results show that the proposed architecture performs better than the legacy \ac{ls} scheme at high speeds. Hence, given their better resilience to high channel variations in vehicular mobility scenarios, the proposed \ac{dl} architecture can be used for channel estimation in C-V2X.

As future work, the following investigations can be carried out

\begin{enumerate}
	\item[1.] The proposed model is only trained on limited channel instantiations and \ac{snr} points. The models is expected to show better performance when trained on a more extensive constellation.
	\item[2] Using Transfer learning to re-train the existing model on real-world data.
	\item[3] Modifying / adding more layers to the existing network to increase its performance.
\end{enumerate}


\bibliographystyle{IEEEtran}
\bibliography{ml_channel_estimation}

\begin{thebibliography}{10}
\providecommand{\url}[1]{#1}
\csname url@samestyle\endcsname
\providecommand{\newblock}{\relax}
\providecommand{\bibinfo}[2]{#2}
\providecommand{\BIBentrySTDinterwordspacing}{\spaceskip=0pt\relax}
\providecommand{\BIBentryALTinterwordstretchfactor}{4}
\providecommand{\BIBentryALTinterwordspacing}{\spaceskip=\fontdimen2\font plus
\BIBentryALTinterwordstretchfactor\fontdimen3\font minus
  \fontdimen4\font\relax}
\providecommand{\BIBforeignlanguage}[2]{{%
\expandafter\ifx\csname l@#1\endcsname\relax
\typeout{** WARNING: IEEEtran.bst: No hyphenation pattern has been}%
\typeout{** loaded for the language `#1'. Using the pattern for}%
\typeout{** the default language instead.}%
\else
\language=\csname l@#1\endcsname
\fi
#2}}
\providecommand{\BIBdecl}{\relax}
\BIBdecl

\bibitem{Hermansson2011}
M.~Hermansson and V.~Skoda, ``{Evaluating channel estimation methods for
  802.11p systems},'' 2011.

\bibitem{Beek1995}
J.-J. Beek, O.~. Edfors, M.~. Sandell, S.~K. Wilson, and P.~O. B{\"{o}}rjesson,
  ``{On channel estimation in OFDM systems},'' vol.~2, pp. 815--819, 1995.

\bibitem{Goodfellow}
I.~Goodfellow, Y.~Bengio, and A.~Courville, \emph{{Deep learning}}.

\bibitem{Jiang2017}
C.~Jiang, H.~Zhang, Y.~Ren, Z.~Han, K.-C. Chen, and L.~Hanzo, ``{Machine
  Learning Paradigms for Next-Generation Wireless Networks},'' \emph{IEEE
  Wireless Communications}, vol.~24, no.~2, pp. 98--105, apr 2017.

\bibitem{Wang2017}
T.~Wang, C.-K. Wen, H.~Wang, F.~Gao, T.~Jiang, and S.~Jin, ``{Deep Learning for
  Wireless Physical Layer: Opportunities and Challenges},'' oct 2017.

\bibitem{Kim2018}
H.~Kim, Y.~Jiang, R.~Rana, S.~Kannan, S.~Oh, and P.~Viswanath, ``{Communication
  Algorithms via Deep Learning},'' may 2018.

\bibitem{OShea2017}
T.~O'Shea and J.~Hoydis, ``{An Introduction to Deep Learning for the Physical
  Layer},'' \emph{IEEE Transactions on Cognitive Communications and
  Networking}, vol.~3, no.~4, pp. 563--575, dec 2017.

\bibitem{Ortuno1992}
I.~Ortuno, M.~Ortuno, and J.~Delgado, ``{Error correcting neural networks for
  channels with Gaussian noise},'' in \emph{[Proceedings 1992] IJCNN
  International Joint Conference on Neural Networks}, vol.~4.\hskip 1em plus
  0.5em minus 0.4em\relax IEEE, 1992, pp. 295--300.

\bibitem{Bruck1989}
J.~Bruck and M.~Blaum, ``{Neural networks, error-correcting codes, and
  polynomials over the binary n-cube},'' \emph{IEEE Transactions on Information
  Theory}, vol.~35, no.~5, pp. 976--987, 1989.

\bibitem{Sattiraju2018}
R.~Sattiraju, A.~Weinand, and H.~D. Schotten, ``{Performance Analysis of Deep
  Learning based on Recurrent Neural Networks for Channel Coding},'' \emph{To
  be published in Proceedings of the IEEE Advances networks and
  Telecommunication Systems (IEEE ANTS)}, nov 2018.

\bibitem{Sattiraju2017a}
R.~Sattiraju, J.~Kochems, and H.~D. Schotten, ``{Machine learning based
  obstacle detection for Automatic Train Pairing},'' in \emph{2017 IEEE 13th
  International Workshop on Factory Communication Systems (WFCS)}.\hskip 1em
  plus 0.5em minus 0.4em\relax IEEE, may 2017, pp. 1--4.

\bibitem{Sattiraju2019}
------, ``{To Supervise or not - ML based UWB Obstacle Detection - Conference
  papers - VDE Publishing House},'' in \emph{Mobilkommunikation –
  Technologien und Anwendungen - 24. ITG-Fachtagung}.\hskip 1em plus 0.5em
  minus 0.4em\relax Osnabrueck, Germany: VDE, 2019, p.~6.

\bibitem{Weinand2017a}
A.~Weinand, M.~Karrenbauer, R.~Sattiraju, and H.~D. Schotten, ``{Application of
  Machine Learning for Channel based Message Authentication in Mission Critical
  Machine Type Communication},'' nov 2017.

\bibitem{Chen1990}
S.~Chen, G.~Gibson, C.~Cowan, and P.~Grant, ``{Adaptive equalization of finite
  non-linear channels using multilayer perceptrons},'' \emph{Signal
  Processing}, vol.~20, no.~2, pp. 107--119, jun 1990.

\bibitem{Sun2006}
\BIBentryALTinterwordspacing
J.~Sun and D.-F. Yuan, ``{Neural Network Channel Estimation Based on Least Mean
  Error Algorithm in the OFDM Systems}.''\hskip 1em plus 0.5em minus
  0.4em\relax Springer, Berlin, Heidelberg, 2006, pp. 706--711. [Online].
  Available: \url{http://link.springer.com/10.1007/11760023{\_}104}
\BIBentrySTDinterwordspacing

\bibitem{Balevi2019}
\BIBentryALTinterwordspacing
E.~Balevi and J.~G. Andrews, ``{Deep Learning-Based Channel Estimation for
  High-Dimensional Signals},'' Tech. Rep., 2019. [Online]. Available:
  \url{https://arxiv.org/pdf/1904.09346.pdf}
\BIBentrySTDinterwordspacing

\bibitem{Ye2018}
\BIBentryALTinterwordspacing
H.~Ye, G.~Y. Li, and B.-H. Juang, ``{Power of Deep Learning for Channel
  Estimation and Signal Detection in OFDM Systems},'' \emph{IEEE Wireless
  Communications Letters}, vol.~7, no.~1, pp. 114--117, feb 2018. [Online].
  Available: \url{http://ieeexplore.ieee.org/document/8052521/}
\BIBentrySTDinterwordspacing

\bibitem{Cheng2016}
\BIBentryALTinterwordspacing
C.-H. Cheng, Y.-H. Huang, and H.-C. Chen, ``{Channel estimation in OFDM systems
  using neural network technology combined with a genetic algorithm},''
  \emph{Soft Computing}, vol.~20, no.~10, pp. 4139--4148, oct 2016. [Online].
  Available: \url{http://link.springer.com/10.1007/s00500-015-1749-7}
\BIBentrySTDinterwordspacing

\bibitem{Soltani}
\BIBentryALTinterwordspacing
M.~Soltani, V.~Pourahmadi, A.~Mirzaei, and H.~Sheikhzadeh, ``{Deep
  Learning-Based Channel Estimation},'' Tech. Rep. [Online]. Available:
  \url{https://github.com/Mehran-Soltani/ChannelNet}
\BIBentrySTDinterwordspacing

\bibitem{Kingma2014}
D.~P. Kingma and J.~Ba, ``{Adam: A Method for Stochastic Optimization},'' dec
  2014.

\end{thebibliography}

\section*{Acknowledgments}
Part of this work has been performed in the framework of the BMVI project Connected Vehicle of the Future (ConVeX). The authors would like to acknowledge the contributions of their colleagues, although the views expressed are those of the authors and do not necessarily represent the project.

\end{document}